# Radiation Hardness of Fiber Bragg Grating Thermometers

Lonnie T. Cumberland, Ryan Fitzgerald, Nikolai N. Klimov, Ronald E. Tosh, Ileana M. Pazos, Zeeshan Ahmed
*Physical Measurement Laboratory, National Institute of Standards and Technology, Gaithersburg, MD, USA*

*Corresponding author:* ryan.fitzgerald@nist.gov

*Photonics sensing has long been valued for its tolerance to harsh environments where traditional sensing technologies fail. As photonic components continue to evolve and find new applications, their tolerance to radiation is emerging as an important line of inquiry. Here we report on our investigation of the impact of gamma-ray exposure on the temperature response of fiber Bragg gratings. At 25 °C, exposures leading to an accumulated dose of up to 600 kGy result in complex dose-dependent drift in Bragg wavelength, significantly increasing the uncertainty in temperature measurements obtained if appreciable dose is delivered over the measurement interval. We note that temperature sensitivity is not severely impacted by the integrated dose, suggesting such devices could be used to measure relative changes in temperature.*

The past two decades have witnessed steady progress in photonics leading to a communications revolution. Tools developed for the telecommunications industry are now being exploited to develop sensors for a wide variety of applications and deployment scenarios.[1-5] Photonic sensors and communication devices are particularly valuable for operation in harsh environments such as those encountered in space exploration and nuclear power plants due to their small size, low power consumption, and tolerance to environmental influences such as mechanical vibrations.[6, 7]

Photonic sensing relies on exploiting the sensitivity of the device to changes in the refractive index of the host material. For example, in a fiber Bragg grating (FBG), the resonant condition is directly proportional to the refractive index of the waveguide. A small change in the refractive index e.g., due to a rise in temperature, leads to a significant change in resonance wavelength which has been exploited for photonic thermometry.[8-10] The sensitivity of photonic devices to small changes in refractive index raises serious questions about the significance of systematic uncertainties that might affect their usability in high radiation environments.[6, 7] Radiation induced damage is known to cause local point defects, dislocations, and formation of color centers, all of which contribute to local changes in refractive index.[11] In principle, these changes significantly degrade a sensor's measurement sensitivity and accuracy.

Previous studies of FBG sensors in radiation environments, with dose rates ranging from a few Gy/h to a few kGy/h, indicate that these devices continue to function over years of exposure to radiation and aggregate dose accumulations exceeding 1 MGy.[12-14] At all dose rates, the optical fibers show increased attenuation with accumulated dose, and some researchers have observed small but significant drifts in Bragg resonances that suggest the resonance wavelength redshifts with increasing dose rate for accumulated doses below about 25 kGy [13], while others have reported a blue shift of comparable magnitude (< 0.1 nm) in studies that included aggregate doses approaching 0.5 MGy [13,5]. Insofar as many industrial irradiation processes deliver doses ranging from 10 Gy to 1 MGy, such dose-related shifts in Bragg wavelength could contribute ≈10 K systematic uncertainty to *in situ* temperature measurements unless a more precise understanding could be obtained of how the shift is correlated with dose and dose rate over the relevant domain of irradiation conditions.

In this study, we present our investigation of FBG thermometers exposed to gamma (γ)-rays. Here we have systematically exposed Ge-doped silica fibers with two commonly used cladding materials – polyimide and ORMOCER (Organic Modified Ceramics)– to varying levels of radiation, leading up to 600 kGy of absorbed dose. We observe a complex dose-dependent change in the Bragg resonance, in which magnitude and direction of wavelength shift vary significantly over the dose range studied. The experimental measurement apparatus has been described in detail elsewhere [8]. Briefly, the FBG sensors, acquired from Micron Optics and FBGS, are probed using a custom-built interrogation system. In this setup, a C-band laser (New Focus TLB-6700 series) is swept over the sensor resonance. Ten percent of the laser power was immediately picked up from the laser output for wavelength monitoring (HighFinesse WS/7; absolute accuracy = 0.16 pm) while the rest, after passing through an optical circulator, was injected in the FBG. Six FBG devices (three with polyimide cladding, labelled FBG 1-3, and three with ORMOCER cladding, labelled FBG 4-6) were exposed to multiple dose fractions of γ-radiation to achieve dose levels ranging from 0.1 Gy to 600 kGy (1 Gy = 100 rad) in the NIST high dose irradiation facility.[15] The absorbed dose to SiO2 in the FBG core was calculated from dose-to-water standards by Monte Carlo simulation of the FBG irradiation geometry.

The impact of dose on resonance peak position is shown in Fig 1 for FBGs 1 through 6, in which wavelength shift vs. absorbed dose is determined relative to the peak position obtained at an absorbed dose of 0.1 kGy. Our results indicate a distinct difference in response of the two groups of FBGs within the initial 0.1 kGy of absorbed dose, with the ORMOCER-coated devices (FBGs 4-6) showing a dramatic redshift of ≈0.75(15) nm, whereas the polyimide-coated fibers (FGBs 1 through 3) exhibit small shifts that are comparable in magnitude to experimental uncertainties (1 standard deviation) as determined by reproducibility of measurements (for which representative error bars are provided on a few of the data points). For absorbed doses above 0.1 kGy, clear differences in response of the two groups of fibers are no longer discernible – all six fibers exhibited a weak tendency to blueshift with absorbed doses between 0.1 kGy and 1 kGy, while all but one showed a tendency to redshift as the accumulated dose grows to 300 kGy. The total shift in peak center for the two fibers that received 300 kGy dose, 135 pm and 165 pm, is the equivalent of ≈13.5 °C and ≈16.5 °C spurious increase in temperature, respectively (i.e. as would be registered by each device in normal thermometric monitoring). However, when FBG 3 was exposed to another 300 kGy dose, for a total integrated dose of 600 kGy, the peak center blue-shifted by 119 pm. The origins of this non-linear dose dependence of the peak center are not presently understood.

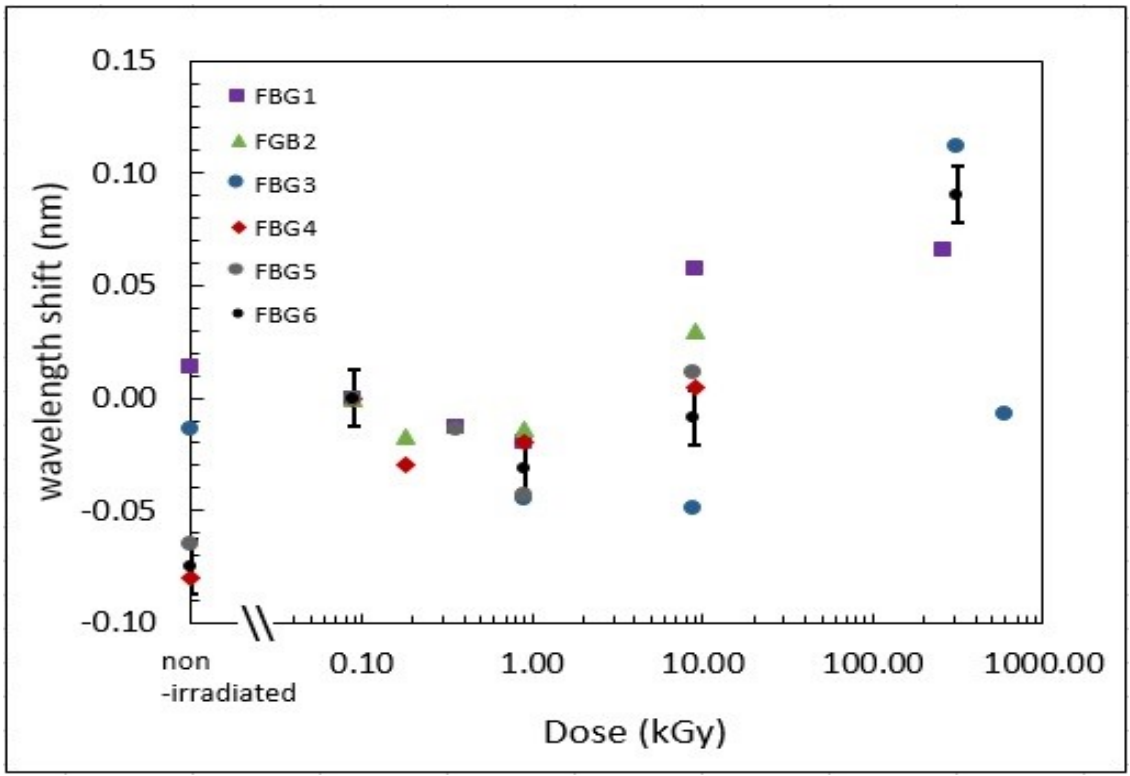

**Figure 1: Dose dependent changes in peak center for FBG sensors show a complex behavior. Uncertainty bars shown represent 1 standard deviation**

We also investigated the effect of radiation dose on temperature sensitivity of three other FBGs by measuring temperature dependence of the resonance wavelength before and after delivering a single dose fraction of 250 kGy. As shown in Fig 2, temperature scans acquired after radiation exposure exhibit a net redshift of ≈120 pm, comparable to what is shown for FBGs 1, 3 and 6 in Figure 1 (i.e., all FBGs in the initial batch of six which had received as much dose), but the temperature sensitivity (slope of plots in Figure 2) of the FBG is not as significantly impacted by radiation dose. Our results therefore indicate that while accumulated dose can result in very large errors in temperature measurements, such devices may still be suitable for relative temperature measurements e.g. tracking temperature rise while a reactor is operational.

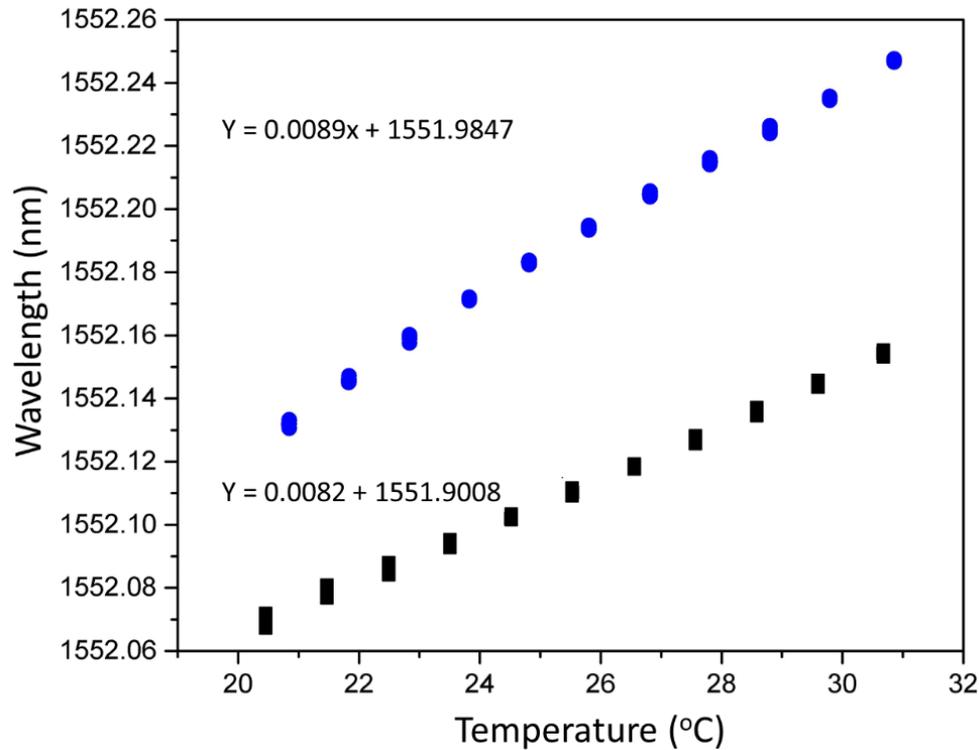

**Figure 2: Temperature sensitivity of a FBG sensor exposed to 250 kGy, 8.0 pm/K, is in range of values expected for unexposed fibers.**

In summary, we have demonstrated that FBG sensors show significant and complex peak center drift due to accumulated dose. The temperature sensitivity of these devices, however, does not show significant changes. Our results are generally consistent with previous results that have reported baseline drifts in FBG resonance due to radiation exposure. [13] Our results quantitatively show the complex behavior is not due to measurement uncertainties in the measurement apparatus, or the outer coating material. The baseline change likely derives from complex changes in the fiber itself. A deeper understanding of dose-dependent changes in optical fibers could nevertheless enable integrated dose measurements over an arbitrary range of absorbed dose with the use of appropriate correction factors.